\documentclass[a4paper,11pt,epsfig,rotating,amssymb,amstex,graphicx,curves]{appolb}
\usepackage{calc}
\usepackage{latexsym}
\usepackage{makeidx}
\usepackage{epsfig}
\usepackage{graphicx}
\usepackage{epstopdf}
\begin{document}
\title{R-hadrons at ATLAS -discovery prospects and properties
\thanks{Presented at Physics at LHC, July 2006, Cracow Poland }%
}
\author{Marianne Johansen
\address{On behalf of the ATLAS Collaboration
\address {
Department of Physics, Stockholm University  \\ 
marianjo@physto.se}}}
\maketitle

\begin{abstract}
R-hadrons are massive, meta-stable particles predicted in several Supersymmetry scenarios. 
Studies exploring the discovery potential of R-hadrons at the ATLAS detector have mainly focused on gluino R-hadrons. These studies have shown that gluino R-hadrons should be discovered in early running of the LHC, that they are easily isolated by simple cuts and that their mass can be measured to an accuracy of a few percent.
\end{abstract}
%\PACS{11.30.Pb, 11.90.+t, 14.80.-j}
\section{Introduction}
Stable Massive Particles (SMPs)\cite{dave} appear in many theories beyond the Standard Model. They are predicted in supersymmetry(SUSY)\cite{susy} models, such as Split-SUSY\cite{splitsusy} and Gauge Mediated Supersymmetry Breaking\cite{gmsb}. They are also predicted in other exotic scenarios, e.g. Universal Extra Dimensions\cite{ued} and lepto-quark theories\cite{leptoquarks}. Because of this both discovery and non discovery is important in excluding different exotic models. 
\par SMPs containing a heavy colored particle are called R-hadrons. ATLAS\cite{atlas} studies focus on gluino R-hadrons that occur in Split-SUSY. Split-SUSY suggests that the hierarchy problem can be addressed by the same fine-tuning mechanism that solves the cosmological constant problem, making low energy Supersymmetry uncalled for. Given this condition, supersymmetry can be broken at a very high energy scale, leading to heavy scalars, light fermions and a light finely tuned Higgs particle\cite{splitsusy}. Within this phenomenological picture squarks will be much heavier than gluinos. he gluino decay will then be suppressed, causing gluinos to be the meta-stable. If the lifetime of the gluinos are long enough they will hadronise giving final states of R-mesons ($\tilde{g}q\bar{q}$), R-baryons ($\tilde{g}qqq$) and so-called R-gluinoballs ($\tilde{g}g$).
\par The details of R-hadron interactions in matter are highly uncertain. However, some features are well understood. The gluino can be regarded as a heavy non-interacting spectator, surrounded by a cloud of interacting quarks\cite{magnus}. R-hadrons change their properties through interaction with the detector, most R-mesons will turn into R-baryons\cite{aafke} and they can also change the sign of their electric charges. 
\par At the Large Hadron Collider (LHC) the gluino R-hadrons are pair produced approximately back-to-back in the transverse plane. PYTHIA\cite{pythia}, with default settings, predicts that 55$\%$ of the R-hadrons are produced neutral and escape detection in the Inner Detector. Charged R-hadrons can leave tracks in both the Inner Detector and the Muon Spectrometer, given that the lifetime is long enough. Furthermore, charge flipping complicates this picture;  R-mesons or R-baryons produced neutral  can become charged and vice versa. However, charge flipping is a powerful tool in isolating a pure sample of R-hadrons. 
\par R-hadrons are heavy and in many cases move with a speed less than the speed of light; their arrival in the Muon Spectrometer will therefore be delayed compared to ordinary muons. R-hadrons with a value of $\beta$ larger than 0.7 are always identified with the same bunch crossing as the rest of the event and satisfy the requirements of the trigger\cite{aafkethesis}. 
\section{ATLAS studies on R-hadrons}
Within the last couple of years (2004-2006) there have been several
studies of R-hadrons at ATLAS. They have all been studies of gluino
R-hadrons within the Split-SUSY scenario, using the implementation of
Kraan's interaction model in GEANT3\cite{aafke}\cite{geant3}. The interaction model was recently implemented\cite{rasmus} in GEANT4\cite{geant4} as well.
\begin{figure}[htb]
\centering
%\vspace{-6 cm}
\includegraphics[width=12cm]{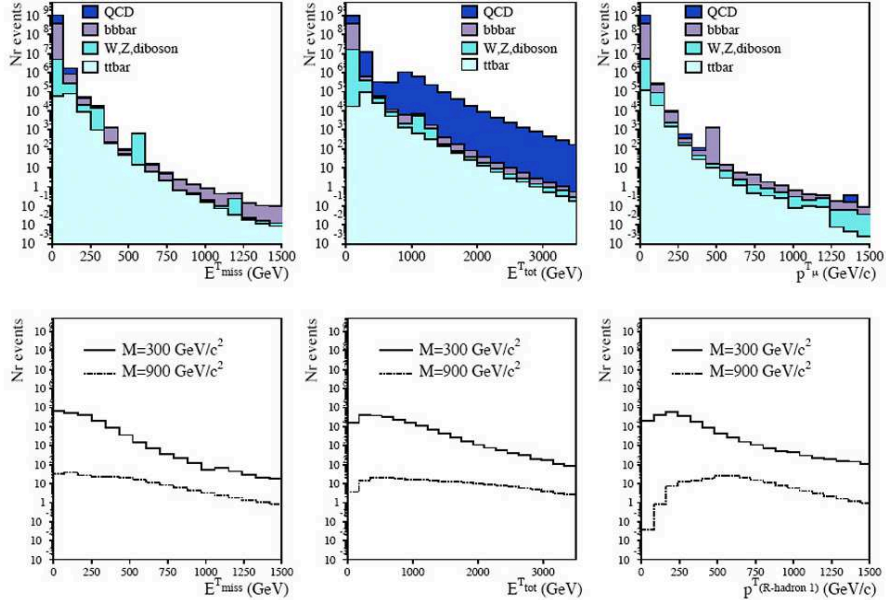}
\caption{\label{fig:aafkefig} The missing transverse energy and the total visible energy, after high level trigger requirements for
background (top) and signal (R-hadron with masses of 300 and 900 GeV/$c^2$). The number of events corresponds to an integrated luminosity of 1 fb$^{-1}$\cite{aafkethesis}}.
\end{figure}
\subsection{Search using global variables}
The discovery of R-hadrons at ATLAS is possible using global variables only \cite{aafke}.
This work used a parametrized detector simulation program (ATLFAST\cite{atlfast}). The focus was to investigate a strategy for discovering R-hadrons at ATLAS solely using global event variables. These were:
\begin{itemize}
\item[$\bullet$] missing transverse energy due to neutral R-hadrons, $E_T^{miss}$.
\item[$\bullet$] the total energy of the event, $E_{TOT}$.
\item[$\bullet$] transverse momentum of the muon track, $p_T$.
\end{itemize}
\par Fig. \ref{fig:aafkefig} shows  $E^{miss}_{T}$,  $E_{tot}$ and $p_{T}$ distributions for background (upper plots) and 300 and 900 GeV/$c^2$ mass R-hadrons  (lower plots). From these plots we see that cutting on the above-mentioned quantities can provide a high discrimination between signal and background.  In fact by using global variables only,  it was shown that R-hadrons up to masses of 1400 GeV/$c^{2}$ could be discovered for an integrated luminosity of 30 fb$^{-1}$. Including time-of-flight information, using the method described in section \ref{sec:mj}, increased this limit up to 1800 GeV/$c^2$. For masses below 1000 GeV/$c^2$ a signal significance above 5 could be reached after a few days. 

\par This strategy is relatively model independent, and shows good
prospects for discovering any Stable Massive Particle, regardless of
scenario.  However it is not suitible for measuring the R-hadron
quantum numbers and can not distinguish between a gluino R-hadron and
another SMP.

\begin{figure}[ht]
    %\vspace{-10cm}
    \includegraphics[width=0.90\linewidth]{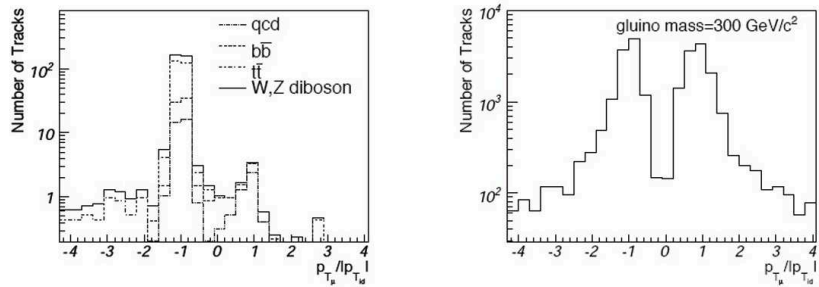}
\caption{\label{fig:magnusfig}Signed $p_T$ in the Muon Spectrometer over the absolute value of $p_T$ in the inner detector for background (left) and signal (right) for particles produced with negative charge.}
\end{figure}
\subsection{Search using charge flippers}
A full GEANT3 detector simulation of R-hadron interactions was used for the first time\cite{magnus}. The purpose of this work was to show how the charge flipping property could be used to isolate a pure sample of R-hadrons. Three different selection criteria were used: 
\begin{itemize}
\item[1.] Opposite charge tracks in Inner Detector (ID) and Muon Spectrometer (MS).
\item[2.] Two same sign, high $p_{T}$-tracks in the Muon Spectrometer.
\item[3.] Two same sign, high $p_{T}$-tracks in MS and explicitly no tracks in ID. 
\end{itemize}
For masses 300 and 500 GeV/$c^{2}$ tracks were required to have $p_{T}>150$ GeV/$c$, while for 1000 GeV/$c^{2}$ masses the requirement on track momenta was increased to $p_{T}>350$ GeV/$c$. Additional requirements were put on the track reconstruction, the quality of  track matching etc. Both signal and QCD, $b\bar{b}$, $t\bar{t}$ and W, Z diboson background were exposed to the same cuts.
\begin{table}
\centering
\begin{tabular}{|cc |c|c|c|c|}
\hline
\multicolumn{2}{|c|}{Source}&\multicolumn{3}{c|}{Number of tracks passed}\\
\multicolumn{2}{|c|}{}&\multicolumn{3}{c|}{Scaled to $2$fb$^{-1}$}\\\cline{3-5}
& & \textbf{\tiny{Selection 1}} & \textbf{\tiny{Selection 2}}&  \textbf{\tiny{Selection 3}}\\
\hline
\hline
Signal&300&21 598 & 16328 &5199  \\
Gluino mass &500 &1742 & 1488 & 455\\
(GeV/$c^2$)&1000& 16&13 & 6 \\
\hline
Background&QCD& 11 &0.9& -\\
$p_{T}>150$ GeV/$c$&$b\bar{b}$& 4.0&1.8 &-\\
&$t\bar{t}$&0.9& 1.8 &-\\
&$W$,$Z$, diboson & 1.1&6.8 &-\\
&total&17& 11.3 &-\\
\hline
Background&QCD& 0.6&-&-\\
$p_{T}>350$ GeV/$c$&$b\bar{b}$& 0.7&-&-\\
&$t\bar{t}$& -&-&-\\
&$W$, $Z$, diboson & 0.33 &0.8&-\\
&total&1.6&0.8&-\\
\hline
\end{tabular}
\caption{\label{fig:magTab}{\bf Selection 1:} Results from selecting opposite charge tracks in Inner Detector and Muon Spectrometer. {\bf Selection 2:} Results from selecting two high $p_{T}$-tracks in the Muon Spectrometer. {\bf Selection 3:} Results from selecting two same sign, high $p_T$ tracks in MS and explicitly no tracks in ID. No background events satisfied the selection. \newline Entries with no number signifies less than one event at 95\% confidence level.}
\end{table}
\par Fig. \ref{fig:magnusfig} shows the distribution of signed transverse momentum ($q \times |p_{T}|$) in  the muon system($p_{T_{\mu}}$) divided by the absolute value of transverse momentum in ID ($p_{T_{ID}}$) for background and signal. Particles in both samples were produced with negative charge. The background (primarily muons from W, Z and top production) distribution in the leftmost plot shows that the SM particles are still negative when detected in the muon system, while approximately 50$\%$ of the R-hadrons have changed electric charge as shown in the rightmost plot.
\par The results from applying these cuts on signal and background events are shown in Table \ref{fig:magTab}. The leftmost column give the result for selection criterion 1, which explicitly selects particles that have changed charge while traversing the detector. As expected a very good background rejection is shown,  since charge flipping does not occur among ordinary muons. 
\par The central column shows the result of picking two same sign high $p_{T}$ tracks in MS. Two same sign muons with large transverse momentum is also a rare signature, and again very few background events satisfy the selection. 
\par The effect of requiring two same sign, high $p_{T}$ tracks in MS and explicitly no track in the Inner Detector is shown in the rightmost column.  R-hadrons that are produced electrically neutral and later convert into charged particles in the calorimeter will be selected. No background events satisfy this selection, leaving a pure sample of R-hadrons.
\par From Table \ref{fig:magTab} we see that selection criteria 1-3 all give very good background rejection. Also note that criterion 1 and 3 can also identify the particles as gluino R-hadrons, since no other R-hadron or SMP can flip electric charge.
\begin{figure}[ht]
\centering
%\vskip -7 cm
\begin{minipage}[ht]{0.45\linewidth}
    \hskip -1 cm
    \includegraphics[height = 6 cm]{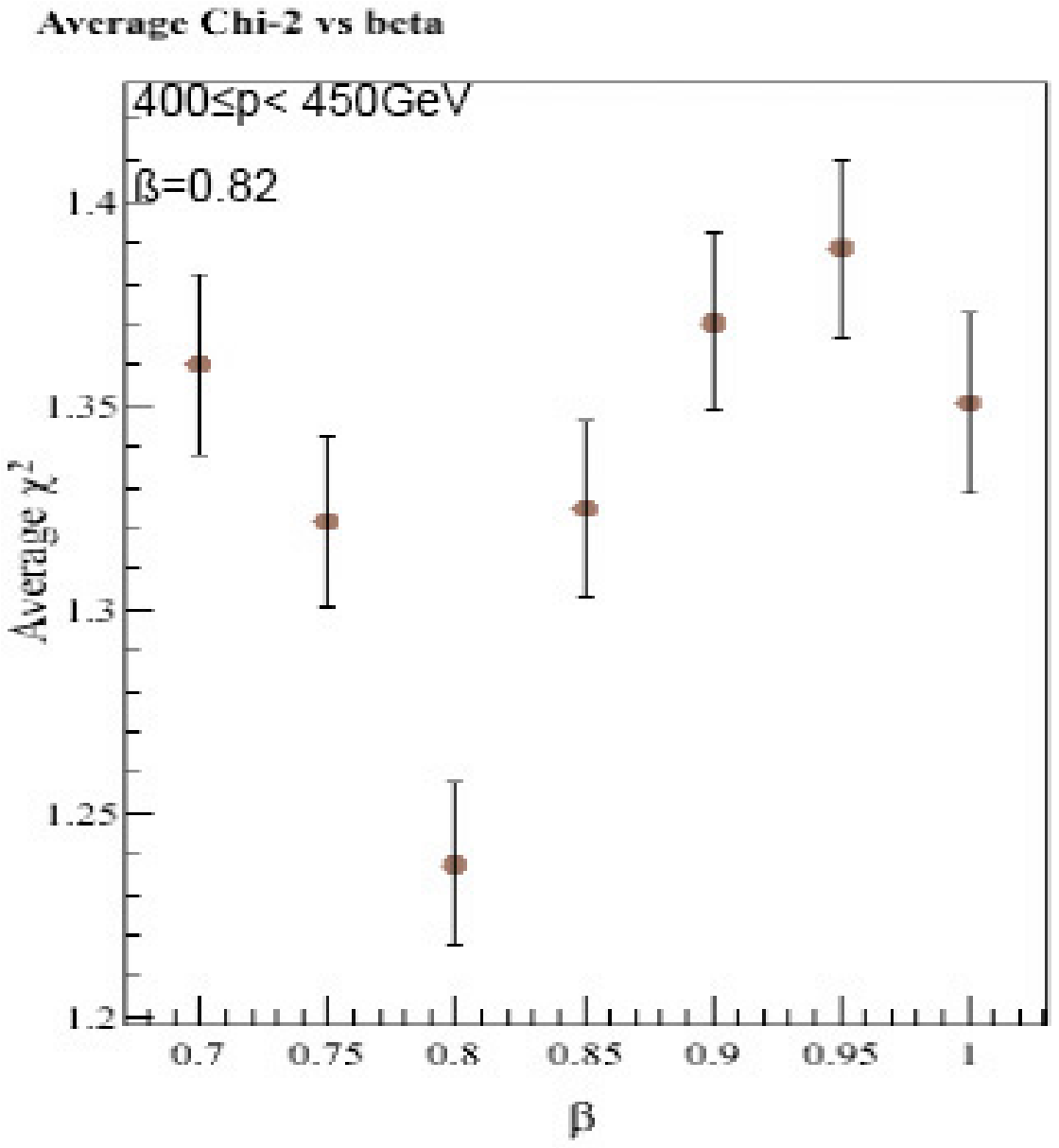}
\end{minipage}
\hskip -1 cm
\begin{minipage}[ht]{0.35\linewidth}
\vskip 4 mm
\includegraphics[height = 5.5 cm]{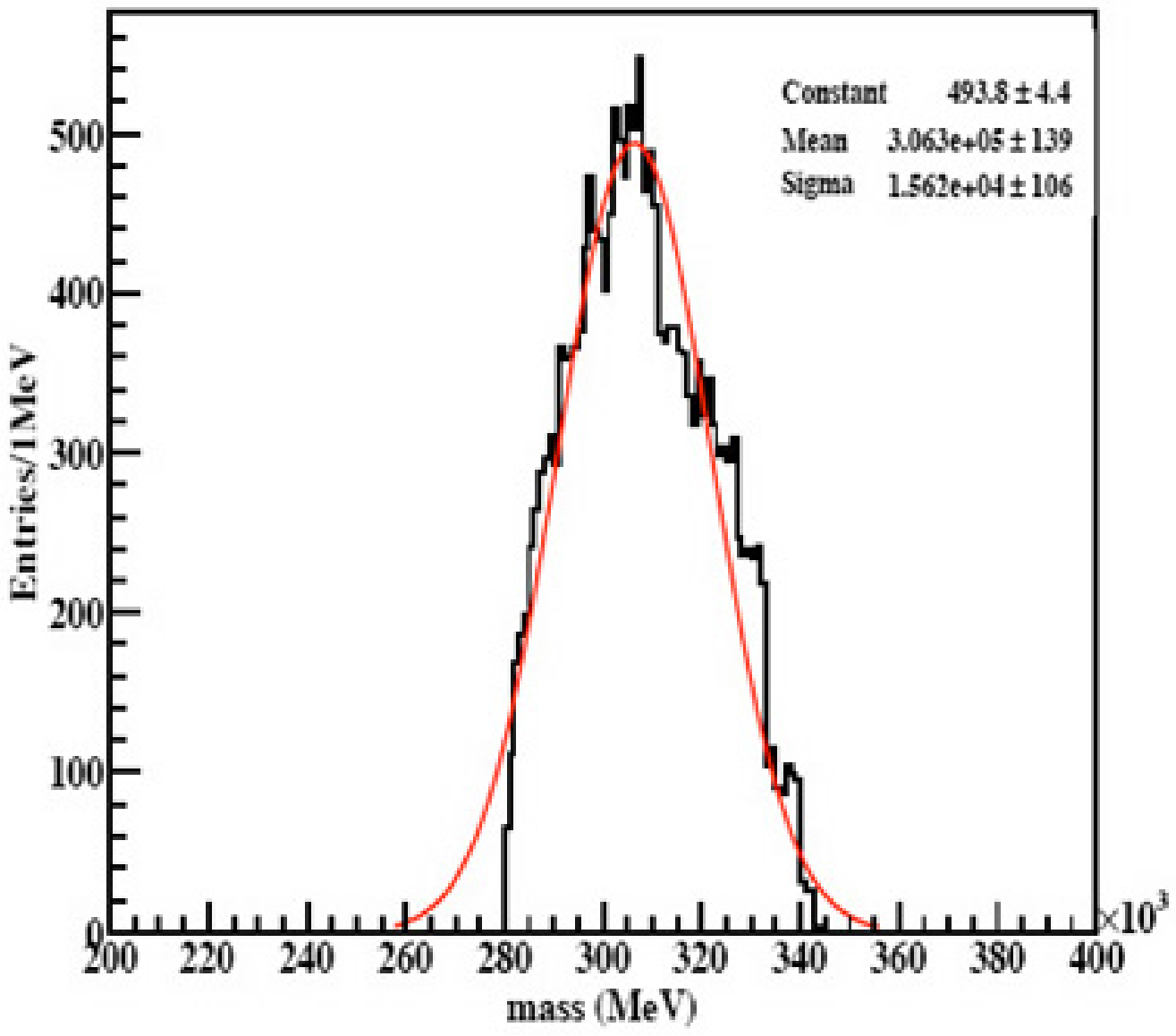}
\end{minipage}
\caption{\label{fig:betaFit}{\bf Left:} Average $\chi^{2}$ as a function of $\beta$. The distribution has a minimum around 0.8 which matches the true value of $\beta=0.82$ {\bf Right:}Reconstructed mass of 300 GeV/$c^{2}$ R-hadrons. }
\end{figure}
\subsection{\label{sec:mj} Mass determination of R-hadrons}
The most recent ATLAS-study\cite{mj}  illustrates the possiblility of measuring the mass of gluino R-hadrons using the time-of-flight method\cite{polesello}. Information on the track fitting parameters is available in the full GEANT3 detector simulation. Here the Moore\cite{moore} track-fitting reconstruction algorithm is tuned for relativistic particles with $\beta=1$. For heavier particles, like  R-hadrons, this will lead to an over-estimation of the drift time\footnote{  (Drift-time = Total time $- L/c\beta$. where L is the distance traveled, c the speed of light and $\beta = \frac{v}{c}$ ).}and will give a bad $\chi^ 2$ for the track  fit. If, however, different values of $\beta$ are assumed during track-reconstruction, the $\chi^ 2$-distribution will have a minimum where the true value of $\beta$ is, as shown in the left plot of Fig.\ref{fig:betaFit}. In this way a measurement of $\beta$ can be obtained. The measured $\beta$ can then be used in the relativistic momentum relation to extract the mass. The plot on the right in Fig. \ref{fig:betaFit} shows the reconsturcted mass for 300 GeV R-hadrons, in this plot the mean value is 306.3 $\pm$ 0.1 GeV/$c^{2}$.
\section{Conclusion}
Since SMPs are predicted in many scenarios beyond the Standard Model, they will have to be searched for at ATLAS. Their discovery or non-discovery will be an important tool in excluding many exotic models. One such scenario is Split-SUSY where gluinos hadronise to form meta-stable massive R-hadrons. Studies in ATLAS have shown that early discovery of R-hadrons up to 1400 GeV/$c^2$ masses is possible, and that the charge-flipping property of gluino R-hadrons can be used to isolate a pure signal sample. This allows for determination of the gluino R-hadron mass up to a few percent.

\end{document}